# The Delay-Insensitivity, the Hazard-Freedom, the Semi-Modularity and the Technical Condition of Good Running of the Discrete Time Asynchronous Automata

Serban E. Vlad

Str. Zimbrului, Nr.3, Bl.PB68, Et.2, Ap.11, 3700, Oradea, Romania
serbanvlad@excite.com, http://www.oradea.ro/serban/index.html

*Abstract* *The paper defines and characterizes the delay-insensitivity, the hazard-freedom, the semi-modularity and the technical condition of good running of the discrete time asynchronous automata.*

## 1. Introduction

The asynchronous automata are the models of the asynchronous circuits. A safety property of such automata states that some bad thing never happens, by 'bad thing' being understood loosing all the information on the behavior of the circuits due to their non-determinism. This property is also considered to be of determinism, since it means the disappearance, in a certain sense, of the uncertainties that characterize the asynchronous citcuits.

The safety properties that we shall refer to in this paper are:

- delay-insensitivity and the technical condition of good running: the existence of certain transitions does not depend on the values of the delays of the circuits that are unknown

- hazard-freedom: delay-insensitivity + the request of monotonous transitions: non monotonous transitions create unpredictable behavior of the circuits

- speed-independence: hazard-freedom under the unbounded gate delay model. Some authors ask imprecisely that the delays after forks be less than the gate delays. Other authors make in this definition the request (of delay-insensitivity) that a unique final class exists, given by the equivalence: two states are equivalent if they are reachable from each other. In the present paper, speed-independence coincides with hazard-freedom

- semi-modularity: if two coordinates are enabled and one switches, the other one is not disabled; this property may be loosened to weak semi-modularity, where for any trajectory (called path here) and any state of the trajectory, the generator function eventually computes coordinatewise the next state.

Most of these notions are well known from the literature, where they are presented informally. Our purpose is to state them by making use of a formalism for the asynchronous automata suggested by a series of papers that Anatoly Chebotarev has published in the 70's and the 80's. Some differences of taxonomy exist relative to these papers.

The assumptions made on the delays of the asynchronous circuits are the following:

*Assumption 1. The delays are concentrated in gates and wires.*

The wires can be explicitly introduced in the description of the circuits as identity elements.

*Assumption 2. The delays are unbounded.*

No lower or upper bounds for the delays are indicated, the only request of this nature is that they be positive. This is a mathematical simplification.

    *Assumption 3. The delays are not constant.*

They are functions varying with temperature, sense of the switch (low-high, respectively high-low), technology and time.

    *Assumption 4. The delays are unknown.*

    In fact during the run of the automaton, as a consequence of assumptions 2,…,4 any delay may be considered to be a sequence of arbitrary positive numbers.

We are grateful to Dr. Anatoly Chebotarev for the great help that he has given during the realization of this work.

## 2. The Autonomous Model

**2.1 Definition** $B_2 = \{0,1\}$ is the binary Boole algebra, endowed with the order $0 \leq 1$ and the discrete topology (where the open sets are all the subsets of $B_2$).

**2.2 Definition** The binary sequence $a_0, a_1, ..., a_k, ...$ is *monotonous*, if it satisfies either of $k \leq k' \Rightarrow a_k \leq a_{k'}$ (*increasing monotonous*), respectively $k \leq k' \Rightarrow a_k \geq a_{k'}$ (*decreasing monotonous*) for all $k, k' \geq 0$.

    The same property takes place for $a_0, a_1, ..., a_k$ a finite family.

**2.3 Definition** The binary sequence $a_0, a_1, ..., a_k, ...$ is *convergent* if it becomes constant starting with a certain rank:
$$\exists \lim_{k \to \infty} a_k \in B_2, \exists N \in N, \forall p \in N, p \geq N \Rightarrow a_p = \lim_{k \to \infty} a_k$$
The number $\lim_{k \to \infty} a_k$ is the *limit of* $a_0, a_1, ..., a_k, ...$ *when* $k$ *tends to infinite*.

**2.4 Remark** $\lim_{k \to \infty} a_k$ is unique because $a_0, a_1, ..., a_k, ...$ are the values of a function $N \to B_2$.

**2.5 Remark** The monotonous sequences are convergent.

**2.6 Definition** The monotony, the convergence and the limit of the $B_2^n$-valued sequences are defined coordinatewise from Definition 2.2 and Definition 2.3. We have from Definition 2.2 the notion of monotony induced coordinatewise for the $B_2^n$-valued finite families too.

**2.7 Definition** The vector $w \in B_2^n$ is called *state*.

**2.8 Definition** The function $g : B_2^n \to B_2^n$ is called *vector field*, or *generator function*.

**2.9 Definition** The coordinate $i$ (or the coordinate function $w_i$), where $i \in \{1,...,n\}$ is *excited* or *enabled* in the state $w$ if $w_i \neq g_i(w)$ and it is *stable* or *disabled* otherwise.

**2.10 Definition** If $w = g(w)$, i.e. if all the coordinates are stable, then $w$ is a *stable state*, or a *point of equilibrium* of the vector field $g$.

**2.11 Definition** The binary relation $m$ on $B_2^n$ is defined by
$$w \, m \, w' \equiv \{i \mid w_i = g_i(w)\} \subset \{i \mid w_i = w_i'\}$$
If $w \, m \, w'$, we say that $w$ *precedes* $w'$ and that $w'$ *follows* $w$.

**2.12 Remark** a) $m$ is reflexive

b) $w\ \boldsymbol{m}\ g(w)$

c) If $w$ is a stable state, then $w\ \boldsymbol{m}\ w' \Rightarrow w = w'$.

d) $w\ \boldsymbol{m}\ w' \wedge w_i \neq w_i' \Rightarrow w_i \neq g_i(w)$

**2.13 Definition** The *reachability* relation $\boldsymbol{M}$ is by definition the transitive closure of $\boldsymbol{m}$:
$$w\boldsymbol{M}w' \equiv \exists w^1,...,\exists w^k, w\ \boldsymbol{m}\ w^1 \wedge ... \wedge w^k\ \boldsymbol{m}\ w'$$
If $w\boldsymbol{M}w'$, we say that $w'$ is *reachable* from $w$.

**2.14 Definition** If $w\boldsymbol{M}w'$, the couple $(w,w')$, usually noted $w \to w'$ is called *transition*, or *transfer* of $w$ in $w'$. We say that $g$ *transfers* $w$ in $w'$.

**2.15 Notation**
$$\boldsymbol{M}(w) = \{w'|\ w\boldsymbol{M}w'\}$$
is the set of the states that are reachable from $w$.

**2.16 Notation**
$$\boldsymbol{M}^{-1}(\tilde{w}) = \{w'|\ w'\boldsymbol{M}\tilde{w}\}$$
is the set of the states from which $\tilde{w}$ is reachable.

**2.17 Remark** The transfers that $g$ makes are called in the literature *non-deterministic*, meaning vaguely that, in general, there exist several $w' \neq w$ with $w\ \boldsymbol{m}\ w'$. In the deterministic situation when a unique such $w'$ exists, $w$ and $g(w) = w'$ differ on exactly one coordinate. On the other hand if $g(w) = w$, then the trivial transition $w \to w$ is considered to be deterministic.

**2.18 Remark** If $w$ is a point of equilibrium of $g$, then from Remark 2.12 c) we have
$$\boldsymbol{M}(w) = \{w\}$$

**2.19 Definition** The sequence $l$ with the terms $w^k \in \boldsymbol{M}(w), k \in \boldsymbol{N}$ is a *path* (or a *trajectory*) with the origin in $w$ if the next conditions are fulfilled:

a) $w^0 = w$

b) $w^0\ \boldsymbol{m}\ w^1 \wedge ... \wedge w^{k-1}\ \boldsymbol{m}\ w^k \wedge ...$

c) $\{i\ |\ \exists a \in \boldsymbol{B}_2, \exists k \geq 0, a = w_i^k = w_i^{k+1} = ... \wedge \overline{a} = g_i(w^k) = g_i(w^{k+1}) = ...\} = \varnothing$

**2.20 Notation** $\qquad \boldsymbol{L}(w) = \{l\ |\ l$ is a path with the origin in $w\}$

**2.21 Remark** a) $\qquad \boldsymbol{M}(w) = \{w^k\ |\ l \in \boldsymbol{L}(w), k \in \boldsymbol{N}\}$

b) $\qquad \{(w',w'')|\ w',w'' \in \boldsymbol{M}(w) \wedge w'\ \boldsymbol{m}\ w''\} = \{(w^k, w^{k+1})\ |\ l \in \boldsymbol{L}(w), k \in \boldsymbol{N}\}$

**2.22 Remark** In this formalism, the *autonomous asynchronous automata* - the taxonomy in [1], [2], [3] is that of *asynchronous circuits without inputs* - are identified with their generator function $g$, $\boldsymbol{N}$ is the time set and the role of $w$ is that of *initial state*. The path $l$ with the origin in $w$ represents the successive values that the state of the automaton takes in discrete time. The condition 2.19 c) means that no coordinate can be excited (without switching) forever.

**2.23 Example** When $g$ defining $\boldsymbol{m}, \boldsymbol{M}, \boldsymbol{M}(w), l, \boldsymbol{L}(w)$ is the constant function $\tilde{g}$ equal to $\tilde{w}$, the new notations are $\tilde{\boldsymbol{m}}, \tilde{\boldsymbol{M}}, \tilde{\boldsymbol{M}}(w), \tilde{l}, \tilde{\boldsymbol{L}}(w)$. For example

$$\tilde{M}(w) = \{w' | \forall i \in \{1,...,n\}, w_i' = w_i \vee w_i' = \tilde{w}_i\}$$

On the other hand, $\tilde{L}(w)$ is described by the fact that the paths $\tilde{l}$ are coordinatewise monotonous and the next inclusion $\{i | w_i^k = \tilde{w}_i\} \subset \{i | w_i^{k+1} = \tilde{w}_i\}$ is true for all $k \in \mathbf{N}$. $\lim_{k \to \infty} w^k$ exists (see Remark 2.5) and it equals $\tilde{w}$. In the special case when $w = \tilde{w}$, $\tilde{l}$ is constant.

**2.24 Proposition** The next statements are equivalent for $l \in L(w)$:

    a) $l$ is convergent

    b) $\exists N \in \mathbf{N}, w^N = g(w^N)$

**Proof** a) $\Rightarrow$ b) $\exists N \in \mathbf{N}$ so that $w^N = w^{N+1} = w^{N+2} = ...$ and we suppose against all reason that $\exists i, w_i^N \neq g_i(w^N)$. 2.19 c) implies the existence of $k \geq 1$ so that $w_i^N \neq w_i^{N+k}$, contradiction with the hypothesis. b) is proved.

b) $\Rightarrow$ a) $\{1,...,n\} = \{i | w_i^N = g_i(w^N)\} \subset \{i | w_i^N = w_i^{N+1}\}$ gives $w^N = w^{N+1}$ etc.

**2.25 Corollary** If $l \in L(w)$ is convergent, then $\lim_{k \to \infty} w^k$ is a stable state.

**2.26 Corollary** If for $l \in L(w)$ some stable state $w^N$ exists, then $l$ is convergent and
$$\lim_{k \to \infty} w^k = w^N$$

**2.27 Corollary**      $\forall w' \in M(w), g(w') = w' \Rightarrow \exists l \in L(w), \lim_{k \to \infty} w^k = w'$

**Proof** From Remark 2.21 a) and Corollary 2.26.

**2.28 Remark** The safety properties from the next sections have also the meaning of giving other points of view on determinism, see Remark 2.17.

## 3. Delay-Insensitivity

**3.1 Notation** If $A(w')$ is a property that depends on $w'$, we use the notation
$$\exists! w', A(w') \equiv \exists w', A(w') \wedge \forall w'', A(w'') \Rightarrow w' = w''$$
for the existence of a unique $w'$ so that $A(w')$.

**3.2 Theorem** The next statements are equivalent:

    a) $\exists \tilde{w}, \forall l \in L(w), \lim_{k \to \infty} w^k = \tilde{w}$

    b) $\begin{cases} \forall l \in L(w), l \text{ is convergent} \\ \exists! \tilde{w}, g(\tilde{w}) = \tilde{w} \wedge M(w) \subset M^{-1}(\tilde{w}) \end{cases}$

    c) $\begin{cases} \forall l \in L(w), l \text{ is convergent} \\ \exists! \tilde{w}, g(\tilde{w}) = \tilde{w} \wedge \tilde{w} \in M(w) \end{cases}$

**Proof** a) $\Rightarrow$ b) Let $l \in L(w)$ arbitrary. It is convergent to a limit $\tilde{w}$ that does not depend on $l$, thus it is unique and it is moreover a stable state, from Corollary 2.25. The terms $w^k$ of $l$ run in all the elements of $M(w)$, when $l \in L(w)$, Remark 2.21 a) and because $w^k M \tilde{w}$, we infer that $M(w) \subset M^{-1}(\tilde{w})$.

b)⇒c) $M(w) \subset M^{-1}(\tilde{w})$ implies $w \, M \, \tilde{w}$ and $\tilde{w} \in M(w)$.

c)⇒a) Because a unique stable state $\tilde{w} \in M(w)$ exists and all $l \in L(w)$ converge, we infer that they converge to the limit $\tilde{w}$.

3.3 **Definition** If one of 3.2 a), b), c) is true, then $g$ is *delay-insensitive* in $w$. We say that $g$ *transfers $w$ in $\tilde{w}$ in a delay-insensitive manner* and that *the transfer $w \to \tilde{w}$ is delay-insensitive*. If $g$ is not delay-insensitive in $w$, it is called *delay-sensitive* in $w$.

3.4 **Remark** (The classification of the situations of delay-sensitivity) $g$ is delay-sensitive in $w$ iff one of the following is true:
    a) $\exists l \in L(w), l$ is not convergent
    b) $\forall w', g(w') = w' \wedge w' \in M(w) \Rightarrow \exists w'', g(w'') = w'' \wedge w'' \in M(w) \wedge w' \neq w''$
These conditions are the negation of 3.2 c), see also 3.1. We get that delay-sensitivity means the existence of an oscillation (case a)), or of several stable states belonging to $M(w)$ (case b)), i.e. the existence of several limits to which the paths $l \in L(w)$ converge, see Corollary 2.27.

3.5 **Notation** In the state transition diagrams to follow, we shall underline the excited coordinates.

3.6 **Example** $g$ is delay-sensitive in $(0,0)$, case 3.4 a):

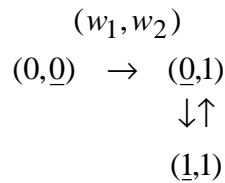

3.7 **Example** $g$ is delay-sensitive in $(0,0)$, case 3.4 b):

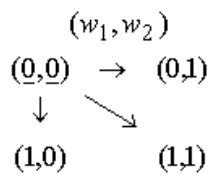

3.8 **Example** $g$ is delay-sensitive in $(0,0)$, case 3.4 a)+b):

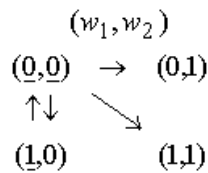

A path $l$ with the origin in $(0,0)$ exists with $w^{2k} = (0,0), w^{2k+1} = (1,0), k \in N$.

3.9 **Example** $g$ is delay-insensitive in $(0,0)$:

$$(w_1, w_2)$$
$$(\underline{0},\underline{0}) \rightarrow (1,0)$$
$$\updownarrow \quad \bowtie \quad \uparrow$$
$$(\underline{0},1) \rightarrow (1,\underline{1})$$

This example contains no path $l$ with the origin in $(0,0)$ so that $w^{2k} = (0,0)$, $w^{2k+1} = (0,1)$, $k \in \mathbf{N}$, due to condition 2.19 c).

## 4. Hazard-Freedom

**4.1 Theorem** The next properties are equivalent

a) $\exists \widetilde{w}, \forall l \in L(w), \lim_{k \to \infty} w^k = \widetilde{w}$ *coordinatewise monotonously*

b) $\begin{cases} \forall l \in L(w), l \text{ is coordinatewise monotonous} \\ \exists! \widetilde{w}, g(\widetilde{w}) = \widetilde{w} \wedge M(w) \subset M^{-1}(\widetilde{w}) \end{cases}$

c) $\begin{cases} \forall l \in L(w), l \text{ is coordinatewise monotonous} \\ \exists! \widetilde{w}, g(\widetilde{w}) = \widetilde{w} \wedge \widetilde{w} \in M(w) \end{cases}$

**Proof** The proof of Theorem 3.2 is repeated, by replacing '$l \in L(w)$ is convergent' with '$l \in L(w)$ is coordinatewise monotonous'.

**4.2 Definition** If one of the equivalent conditions 4.1 a), b), c) is fulfilled, $g$ is called *hazard-free in $w$*. We use to say that $g$ *transfers $w$ in $\widetilde{w}$ in a hazard-free manner* and the transition $w \to \widetilde{w}$ is called *hazard-free*. If $g$ is not hazard-free in $w$, we say that it is *hazardous in $w$*.

**4.3 Remark** Hazard-freedom is obviously stronger than delay-insensitivity, because it asks that the delay-insensitive transition $w \to \widetilde{w}$ be coordinatewise monotonous.

**4.4 Remark** (The classification of the hazards) $g$ is hazardous in $w$ iff at least one of the next statements is true:

a) $\exists l \in L(w), \exists i \in \{1,...,n\}, w_i^0, w_i^1, w_i^2,...$ is not monotonous

b) $\forall w', g(w') = w' \wedge w' \in M(w) \Rightarrow \exists w'', g(w'') = w'' \wedge w'' \in M(w) \wedge w' \neq w''$

**4.5 Remark** We have from Remark 4.4 the possibility that $g$ be delay-insensitive and hazardous in $w$: a) is true with all $l \in L(w)$ convergent to a same limit $\widetilde{w}$ and b) is false. This is the situation from Example 3.9.

**4.6 Definition** If $g$ is delay-insensitive in $w$:
$$\exists \widetilde{w}, \forall l \in L(w), \lim_{k \to \infty} w^k = \widetilde{w}$$
and hazardous in $w$, then the transition $w \to \widetilde{w}$ is called *hazardous*.

**4.7 Remark** Similarly to the conditions 3.2 a) and 4.1 a) where $\exists \widetilde{w}$ meant $\exists! \widetilde{w}$, in the following statements that are supposed to be valid
$$\exists \widetilde{w}, \forall w' \in M(w), g(w') = \widetilde{w}$$
$$\exists \widetilde{w}, L(w) \subset \widetilde{L}(w)$$
$\widetilde{w}$ is the unique state with this property.

**4.8 Remark** Any of the next equivalent conditions

a) $\exists \widetilde{w}, \forall w' \in M(w), g(w') = \widetilde{w}$

b) $\exists \tilde{w}, L(w) \subset \tilde{L}(w)$

implies the hazard-freedom of $g$ in $w$. For example the constant vector fields are hazard-free in any state.

**4.9 Definition** If one of 4.8 a), b) is true, $g$ is *trivially hazard-free in* $w$.

**4.10 Example** $g$ is trivially hazard-free in $(0,0)$:

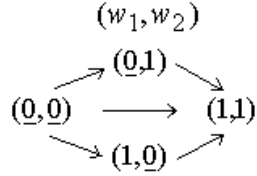

**4.11 Example** $g$ is hazard-free in $w$:

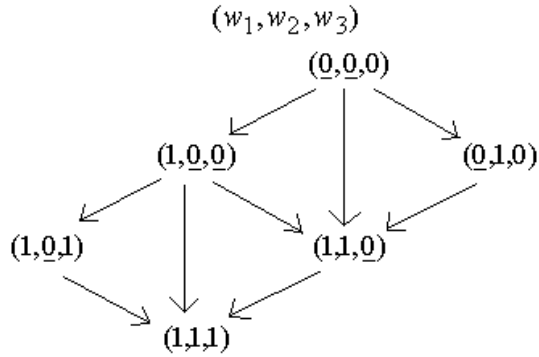

## 5. Semi-Modularity

**5.1 Proposition** The next statements are equivalent:

    a) $\forall w', w'' \in M(w), \forall i \in \{1,...,n\}, w_i' \neq g_i(w') \wedge w'\, \boldsymbol{m}\, w'' \wedge w_i' = w_i'' \Rightarrow w_i'' \neq g_i(w'')$

    b) $\forall w', w'' \in M(w), \forall i, j \in \{1,...,n\}$,

        $w_i' \neq g_i(w') \wedge w'\, \boldsymbol{m}\, w'' \wedge w_i' = w_i'' \wedge w_j' \neq g_j(w') \wedge w_j' \neq w_j'' \Rightarrow w_i'' \neq g_i(w'')$

    c) $\forall l \in L(w), \forall i \in \{1,...,n\}, \forall k \geq 0, w_i^k \neq g_i(w^k) \wedge w_i^k = w_i^{k+1} \Rightarrow w_i^{k+1} \neq g_i(w^{k+1})$

**Proof** a) $\Rightarrow$ b) We suppose that $w', w'' \in M(w)$ and $i, j \in \{1,...,n\}$ are arbitrary with

$$w_i' \neq g_i(w') \wedge w'\, \boldsymbol{m}\, w'' \wedge w_i' = w_i'' \wedge w_j' \neq g_j(w') \wedge w_j' \neq w_j''$$

resulting that

$$w_i' \neq g_i(w') \wedge w'\, \boldsymbol{m}\, w'' \wedge w_i' = w_i''$$

is true and from a) we have

$$w_i'' \neq g_i(w'')$$

b) $\Rightarrow$ a) We suppose that $w', w'' \in M(w)$ and $i \in \{1,...,n\}$ are arbitrary with

$$w_i' \neq g_i(w') \wedge w'\, \boldsymbol{m}\, w'' \wedge w_i' = w_i''$$

If $w' = w''$ then $w_i'' \neq g_i(w'')$ and the proof is complete, so that we may suppose that $w' \neq w''$. Because $w_i' = w_i''$, some $j \neq i$ exists with $w_j' \neq w_j''$ and the definition of $m$ implies $w_j' \neq g_j(w')$. We use the truth of

$$w_i' \neq g_i(w') \wedge w' \, m \, w'' \wedge w_i' = w_i'' \wedge w_j' \neq g_j(w') \wedge w_j' \neq w_j''$$

and b) gives

$$w_i'' \neq g_i(w'')$$

a) $\Leftrightarrow$ c) follows from Remark 2.21 b).

**5.2 Definition** $g$ is *semi-modular in $w$* if one of 5.1 a), b), c) is satisfied.

**5.3 Remark** The taxonomy of semi-modularity is related to lattice theory. It states that an excited coordinate remains excited at least until it switches. It also states that if two coordinates are enabled and one switches, the other one is not disabled.

**5.4 Example** $g$ is semi-modular in $(0,0)$:

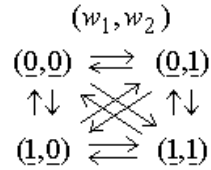

**5.5 Definition** $g$ is *weakly semi-modular in $w$* if:

$$\forall l \in L(w), \forall i \in \{1,...,n\}, \forall k \geq 0, \exists k' \geq k, w_i^{k'} = g_i(w^k)$$

**5.6 Remark** The weak semi-modularity of $g$ in $w$ means that for any path with the origin in $w$, any coordinate $i$ and any state $w^k$, $g$ eventually computes $g_i(w^k)$.

**5.7 Proposition** If $g$ is semi-modular in $w$, then it is also weakly semi-modular in $w$.

**Proof** Let $l \in L(w), i \in \{1,...,n\}$ and $k \geq 0$. If $w_i^k = g_i(w^k)$, then the proposition is proved with $k' = k$ so that we shall suppose in the rest of the proof that $w_i^k \neq g_i(w^k)$.

Step 1 a) $w_i^k \neq w_i^{k+1}$ ; then the proposition is proved with $k' = k+1$

b) $w_i^k = w_i^{k+1}$ ; then the semi-modularity of $g$ in $w$ shows that

$$g_i(w^k) = g_i(w^{k+1}) \neq w_i^k = w_i^{k+1}$$

and we go to

Step 2 a) $w_i^{k+1} \neq w_i^{k+2}$ ; then the proposition is proved with $k' = k+2$

b) $w_i^{k+1} = w_i^{k+2}$ ; then the semi-modularity of $g$ in $w$ shows that

$$g_i(w^k) = g_i(w^{k+1}) = g_i(w^{k+2}) \neq w_i^k = w_i^{k+1} = w_i^{k+2}$$

and we go to
Step 3 …

The condition 2.19 c) shows the existence of some $p \geq 0$ so that

$$w_i^{k+p} \neq w_i^{k+p+1}$$

and the proposition is proved with $k' = k + p + 1$.

**5.8 Example** $g$ is delay-insensitive in $(0,0,0)$ and the transition $(0,0,0) \rightarrow (1,1,0)$ is hazardous

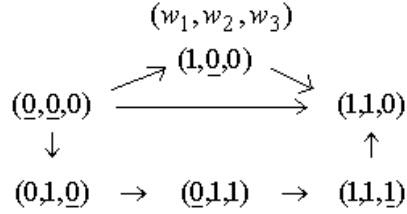

$g$ is weakly semi-modular in $(0,0,0)$ but it is not semi-modular in $(0,0,0)$ because $w_1$ is not excited in $(0,1,0)$.

**5.9 Example** $g$ is hazard-free in $(0,0,0)$ but not semi-modular in $(0,0,0)$, because $w_2$ is not excited in $(1,0,0)$

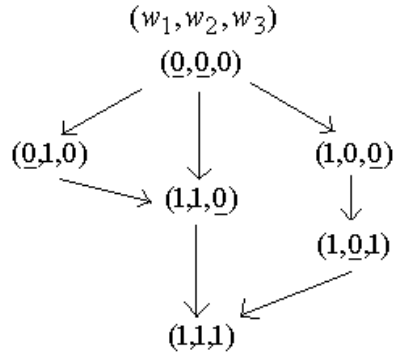

**5.10 Proposition** If $g$ is trivially hazard-free in $w$, then it is semi-modular in $w$.
**Proof** Let $w', w'' \in M(w)$ and $i \in \{1,...,n\}$ with
$$w_i' \neq g_i(w') \wedge w' \, \boldsymbol{m} \, w'' \wedge w_i' = w_i''$$
The hypothesis states that
$$g_i(w') = g_i(w'') = \tilde{w}_i$$
resulting
$$w_i'' \neq g_i(w'')$$

**5.11 Proposition** If $g$ is hazard-free in $w$, then it is weakly semi-modular in $w$.
**Proof** We suppose against all reason that $g$ is hazard-free in $w$ and it is not weakly semi-modular in $w$ and let $l \in L(w), i \in \{1,...,n\}, k \geq 0$ so that
$$\forall k' \geq k, w_i^{k'} \neq g_i(w^k)$$
resulting
$$\forall k' \geq k, w_i^{k'} = w_i^k = \tilde{w}_i \neq g_i(w^k)$$
$l' \in L(w)$ exists with

$$w_i^{'p} = \begin{cases} w_i^p, p = \overline{0,k} \\ \overline{w_i^p}, p = k+1 \end{cases}$$

implying the existence of $k' > k+1$ so that

$$w_i^{'k'} = \tilde{w}_i$$

The sequence $w_i^{'0}, w_i^{'1}, ..., w_i^{'k}, w_i^{'k+1}, ..., w_i^{'k'}, ...$ is not monotonous, contradiction.

## 6. The Technical Condition of Good Running

**6.1 Definition** The *iteratives* of $g$ are the functions $g^j : \mathbf{B}_2^n \to \mathbf{B}_2^n, j \geq 0$ defined in the next manner:

$$g^0(w) = w$$
$$g^{j+1}(w) = g(g^j(w))$$

**6.2 Theorem** The next statements are equivalent:
   a) $\forall w', w'' \in \mathbf{M}(w), w' \mathbf{m} w'' \wedge w'' \neq g(w') \Rightarrow g(w') = g(w'')$
   b) We are in one of the next exclusive situations:
   b.1) $g(w) = w$
   b.2) $\exists p \geq 1, g(w) \neq w \wedge g^2(w) \neq g(w) \wedge ... \wedge g^p(w) \neq g^{p-1}(w) \wedge g^p(w) = g^{p+1} = ...$
   $\forall j \in \{0, ..., p-1\}, \forall w' \in \mathbf{M}(g^j(w)) \wedge \mathbf{M}^{-1}(g^{j+1}(w)) - \{g^{j+1}(w)\}, g(w') = g^{j+1}(w)$
   b.3) $g(w) \neq w \wedge g^2(w) \neq g(w) \wedge ... \wedge g^{p+1}(w) \neq g^p(w) \wedge ...$
   $\forall j \in \mathbf{N}, \forall w' \in \mathbf{M}(g^j(w)) \wedge \mathbf{M}^{-1}(g^{j+1}(w)) - \{g^{j+1}(w)\}, g(w') = g^{j+1}(w)$
   c) For any $l \in \mathbf{L}(w)$, the numbers $k_0, k_1, k_2, ...$ exist so that
$$0 = k_0 < k_1 < k_2 < ...$$
$$w^{k_j} = g^j(w), j \in \mathbf{N}$$

being satisfied one of the next exclusive conditions:
   c.1) $w^{k_0} = w^{k_1} = w^{k_2} = ...$
   c.2) $\exists p \geq 1, w^{k_1} \neq w^{k_0} \wedge w^{k_2} \neq w^{k_1} \wedge ... \wedge w^{k_p} \neq w^{k_{p-1}} \wedge w^{k_p} = w^{k_{p+1}} = ...$
   $\forall j \in \{0, ..., p-1\}, g(w^{k_j}) = g(w^{k_j+1}) = ... = g(w^{k_{j+1}-1}) = w^{k_{j+1}}$
   c.3) $w^{k_1} \neq w^{k_0} \wedge w^{k_2} \neq w^{k_1} \wedge ... \wedge w^{k_{p+1}} \neq w^{k_p} \wedge ...$
   $\forall j \in \mathbf{N}, g(w^{k_j}) = g(w^{k_j+1}) = ... = g(w^{k_{j+1}-1}) = w^{k_{j+1}}$

**Proof** The transitions $w^{k_j} \to w^{k_j+1}$ take place similarly to trivial hazard-freedom in $w^{k_j}$, as $g$ is constant on $\mathbf{M}(g^j(w)) \wedge \mathbf{M}^{-1}(g^{j+1}(w)) - \{g^{j+1}(w)\}$ whenever these sets are non-empty, $j \in \mathbf{N}$.

**6.3 Definition** If one of the conditions 6.2 a), b), c) is fulfilled, then $g$ satisfies the *technical condition of good running* TCGR *in* $w$.

**6.4 Remark** If $g$ satisfies TCGR in $w$ then the transitions $w^{k_j} \to w^{k_j+1}$ are coordinatewise monotonous, $j \in \mathbf{N}$.

**6.5 Example** $g$ satisfies TCGR in $(0,0,0)$:

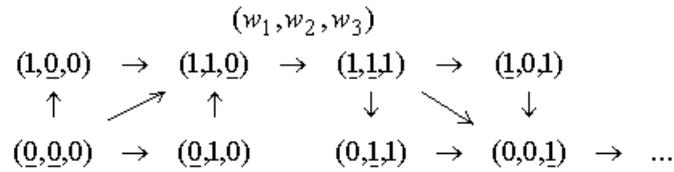

**6.6 Example** $g$ satisfies TCGR in $(0,0)$:

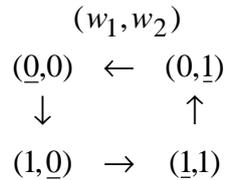

**6.7 Remark** A special case of TCGR in $w$ consists in the situation when $\forall j \in N, g^{j+1}(w)$ and $g^j(w)$ differ on at most one coordinate, like in the previous example, because the sets $M(g^j(w)) \wedge M^{-1}(g^{j+1}(w)) - \{g^{j+1}(w)\}$ contain at most the points $\{g^j(w)\}$. This special case has been called by Grigore Moisil in his pioneering works in automata theory the technical condition of good running from where, by generalization, our taxonomy has resulted.

This special case is called in more recent works the *single bit change* and it is offered as a type of race-free encoding [5].

**6.8 Remark** on the pseudo-periodicity of the paths $l \in L(w)$.

a) We suppose that $g$ satisfies TCGR in $w$. Then $j \geq 0$ and $p \geq 1$ exist so that

$$g^j(w) = g^{j+p}(w)$$

because $B_2^n$ is finite and in the following we fix the least $j, p$ with this property.

b) Let $l \in L(w)$ arbitrary and fixed. The indexes $k_0, k_1, k_2, \ldots$ are defined like at 6.2 c) by:

$$0 = k_0 < k_1 < k_2 < \ldots$$
$$w^{k_m} = g^m(w), m \in N$$

c) The existence of the sets

$$T_l = \{w^k \mid k = \overline{k_0, k_j - 1}\}$$
$$P_l = \{w^k \mid k \geq k_j\}$$

- for which

$$j = 0 \Rightarrow T_l = \varnothing$$
$$p = 1 \Rightarrow P_l = \{w^{k_j}\}$$

are true - shows that $l$ has two periods, of transient regime respectively of permanent regime.

d) The equation

$$w^{k_j + m} = w^{k_{j+p} + m}, m \in N$$

showing the pseudo-periodicity of $l$ during the permanent regime is not generally satisfied.

e) Counterexample $g$ satisfies TCGR in $0$:

$$\underline{0} \leftrightarrows \underline{1}$$

$j = 0, p = 2$ and $l$ is the sequence $0,1,1,0,0,0,1,1,1,1,...$.

**6.9 Proposition** If $g$ satisfies TCGR in $w$ and
$$\exists p \geq 1, g^{p+1}(w) = g^p(w)$$
then $g$ is delay-insensitive in $w$ and
$$\forall l \in L(w), \lim_{k \to \infty} w^k = g^p(w)$$

**6.10 Proposition** The next statements are equivalent:
a) $\begin{cases} g \text{ satisfies TCGR in } w \\ g(w) = w \vee g^2(w) = g(w) \end{cases}$
b) $g$ is trivially hazard-free in $w$

**6.11 Proposition** If $g$ satisfies TCGR in $w$, then it is semi-modular in $w$.
**Proof** Let $w', w'' \in M(w)$ and $i \in \{1,...,n\}$ so that
$$w'_i \neq g_i(w') \wedge w' \, m \, w'' \wedge w'_i = w''_i$$
resulting $w'' \neq g(w')$. TCGR gives $g(w') = g(w'')$ and we infer that
$$w''_i \neq g_i(w'')$$

## 7. The Non-Autonomous Case

**7.1 Remark** We shall identify the spaces $B_2^{n+m}$ and $B_2^n \times B_2^m$.

**7.2 Definition** The vector $z = (w, v) \in B_2^n \times B_2^m$ is called *extended state*, or *total state*. The vector $w$ of the first $n$ coordinates of $z$ is called *state* and the vector $v$ of the last $m$ coordinates of $z$ is called *input*, or *control*.

**7.3 Notation** We shall note $z' = (w', v'), z'' = (w'', v''), \tilde{z} = (\tilde{w}, \tilde{v}),...$

**7.4 Definition** The function $f : B_2^n \times B_2^m \to B_2^n, B_2^n \times B_2^m \ni (w, v) \mapsto f(w, v) \in B_2^n$ is called *vector field with one parameter*, or *generator function with one parameter*. $v$ is the *parameter* of $f$.

**7.5 Notation** Let $\tilde{v} \in B_2^m$ and we note with $f^{\tilde{v}} : B_2^n \times B_2^m \to B_2^n \times B_2^m$ the function
$$f_j^{\tilde{v}}(w, v) = \begin{cases} f_j(w, v), j = \overline{1, n} \\ \tilde{v}_{j-n}, j = \overline{n+1, n+m} \end{cases}$$

**7.6 Remark** From now we shall replace $g$ with $f^{\tilde{v}}$ and we shall make use of the fact that the last $m$ coordinates of $f^{\tilde{v}}$ are the constant functions.

**7.7 Notation** We note with $m^{\tilde{v}}, M^{\tilde{v}}(z), M^{\tilde{v}^{-1}}(z), l^{\tilde{v}}, L^{\tilde{v}}(z)$ the notions resulting from $m, M(w), M^{-1}(w), l, L(w)$ when $g$ is replaced by $f^{\tilde{v}}$.

**7.8 Remark** If $v = \tilde{v}$ or if, more general, the vector field $g$ exists so that
$$\forall z' \in M^{\tilde{v}}(z), f^{\tilde{v}}(z') = g(w')$$
then we get a situation that is equivalent to the autonomous automata and we have
$$M(w) = \{w' | z' \in M^{\tilde{v}}(z)\}$$

**7.9 Definition** If the next condition is fulfilled
$$\exists \tilde{w}, \forall l^{\tilde{v}} \in L^{\tilde{v}}(z), \lim_{k \to \infty} w^k = \tilde{w}$$
then $f^{\tilde{v}}$ is *delay-insensitive in* $z$, otherwise $f^{\tilde{v}}$ is *delay-sensitive in* $z$.

**7.10 Example** $f^1$ is delay-insensitive in $(0,0)$ :

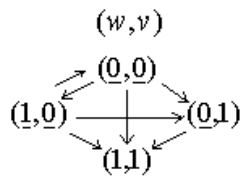

**7.11 Definition** If $f^{\tilde{v}}$ is delay-insensitive in $z$ and moreover if
$$f(z) = w$$
then $f^{\tilde{v}}$ is *delay-insensitive* in $z$ *in the fundamental mode*. We say that $f^{\tilde{v}}$ transfers $z$ in $\tilde{z}$ in a delay-insensitive manner and that *the transfer* $z \to \tilde{z}$ *is delay-insensitive in the fundamental mode*.

**7.12 Remark** If the condition 7.11 of stability in $z$ is true, then
$$z \, m^{\tilde{v}} \, z' \Rightarrow w = w'$$
i.e. the first non-trivial transition of each path $l^{\tilde{v}} \in L^{\tilde{v}}(z)$ takes place on one of the coordinates of the input.

**7.13 Remark** The fact that $f^{\tilde{v}}$ is delay-insensitive in $z$ in the fundamental mode is related to two conditions of stability, in $z$ and in $\tilde{z}$. This is a request of inertial compatibility between the automaton and its input (= the environment).

**7.14 Example** (to be compared with 3.9) $f^0$ is delay-insensitive in $(0,0,1)$ in the fundamental mode:

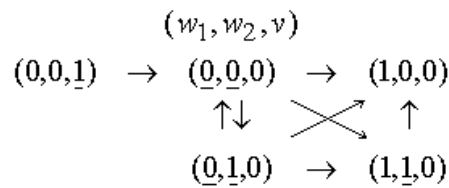

The input is the button that, when pushed, makes the automaton start running.

**7.15 Definition** If $f^{\tilde{v}}$ is delay-insensitive in $z$ and if
$$\forall l^{\tilde{v}} \in L^{\tilde{v}}(z), \forall i \in \{1,...,n\}, w_i^0, w_i^1, w_i^2,... \text{ is monotonous}$$
then $f^{\tilde{v}}$ is *hazard-free in $z$*. If moreover

a) $\exists \tilde{w}, \forall z' \in M^{\tilde{v}}(z), f(z') = \tilde{w}$, then $f^{\tilde{v}}$ is *trivially hazard-free in $z$*

b) $f(z) = w$, then $f^{\tilde{v}}$ is *hazard-free in $z$ in the fundamental mode*.

**7.16 Remark** In the hypothesis that $f^{\tilde{v}}$ is trivially hazard-free in $z$ in the fundamental mode, we have
$$\forall z' \in M^{\tilde{v}}(z), f(z') = w$$
$$\forall l^{\tilde{v}} \in L^{\tilde{v}}(z), \forall k \geq 0, z^k \neq z^{k+1} \Rightarrow v^k \neq v^{k+1}$$

**7.17 Examples** $f^1$ is hazard-free in $(0,0,0)$, non-trivially and not in the fundamental mode at a); $f^1$ is trivially hazard-free in $(0,0)$ but not in the fundamental mode at b); $f^1$ is non-trivially hazard-free in $(0,0)$ in the fundamental mode at c); and $f^1$ is trivially hazard-free in $(0,0)$ in the fundamental mode at d).

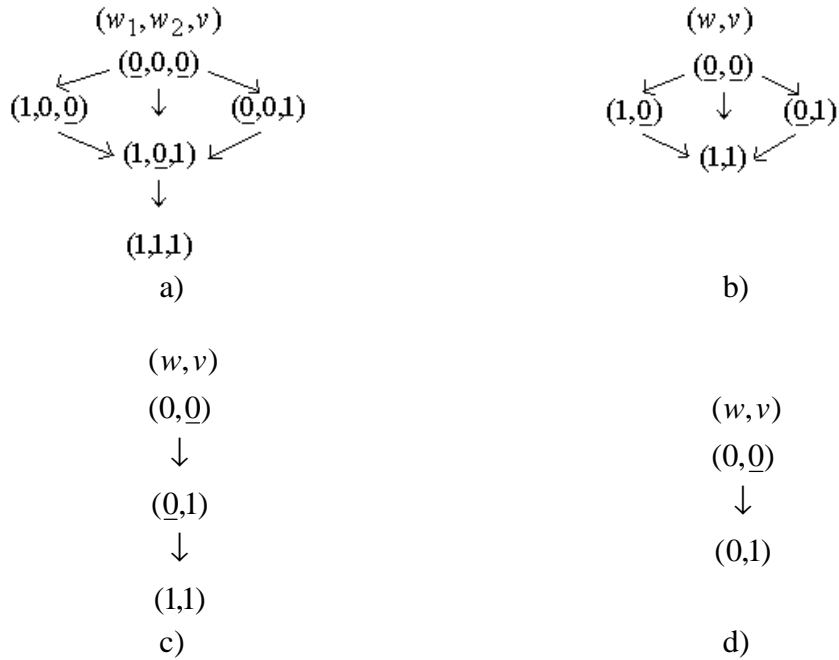

**7.18 Definition** If $f^{\tilde{v}}$ fulfills the condition
$$\forall z', z'' \in M^{\tilde{v}}(z), \forall i \in \{1,...,n\}, w_i' \neq f_i(z') \wedge z' \, m^{\tilde{v}} \, z'' \wedge w_i' = w_i'' \Rightarrow w_i'' \neq f_i(z'')$$
then it is *semi-modular in $z$*. If, moreover, $f(z) = w$ is true, then we say that $f^{\tilde{v}}$ is *semi-modular in $z$ in the fundamental mode*.

**7.19 Examples** $f^1$ is semi-modular in $(0,0)$ but not in the fundamental mode at a):

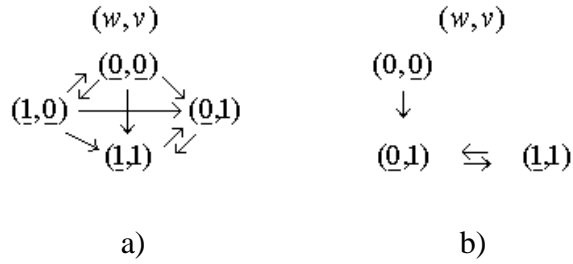

a)                                       b)

and $f^1$ is semi-modular in $(0,0)$ in the fundamental mode at b).

**7.20 Definition** $f^{\tilde{v}}$ is *weakly semi-modular in $z$* if
$$\forall l^{\tilde{v}} \in L^{\tilde{v}}(z), \forall i \in \{1,...,n\}, \forall k \geq 0, \exists k' \geq k, w_i^{k'} = f_i(z^k)$$

By supposing moreover that $f(z) = w$ is satisfied, we say that $f^{\tilde{v}}$ is *weakly semi-modular in $z$ in the fundamental mode*.

**7.21 Example** 5.8 gives the next example, where $f^1$ is weakly semi-modular in $(0,0,0,0)$ in the fundamental mode, but it is not semi-modular in $(0,0,0,0)$:

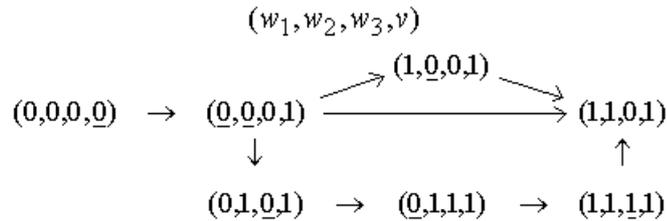

**7.22 Definition** If $f$ fulfills
$$\forall z', z'' \in M^{\tilde{v}}(z), z' m^{\tilde{v}} z'' \wedge w'' \neq f(z') \Rightarrow f(z') = f(z'')$$

or any of the equivalent conditions derived from 6.2 then $f^{\tilde{v}}$ satisfies the *technical condition of good running* TCGR *in $z$*. If moreover $f(z) = w$, then $f^{\tilde{v}}$ satisfies TCGR in $z$ in the *fundamental mode*.

**7.23 Example** We have at 7.19 b) that $f^1$ satisfies TCGR in $(0,0)$ in the fundamental mode. This is a trivial example, the single bit change case.

## 8. Conclusions

The asynchronous automata are governed by non-determinism due to the gate and the wire delays that are unknown. Under certain assumptions, we have defined some situations of safetiness and determinism that are summarized in the next drawing:

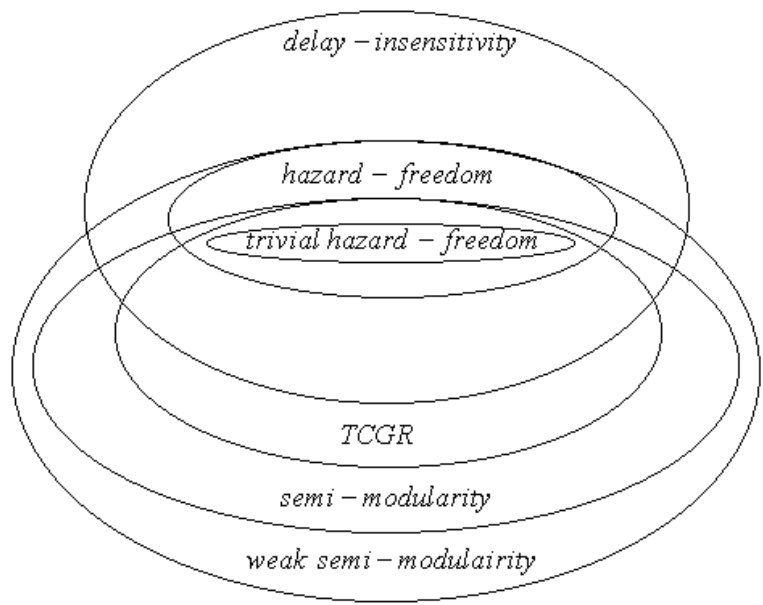